\documentclass[twocolumn,showpacs,preprintnumbers,amsmath,amssymb]{revtex4}

\usepackage{graphicx}
\usepackage{dcolumn}
\usepackage{bm}
\usepackage{verbatim}

\begin{document}

\title{An upper limit on electron antineutrino mass from Troitsk experiment}
\author{V.N.~Aseev, A.I.~Belesev, A.I.~Berlev, E.V.~Geraskin, A.A.~Golubev, 
N.A.~Likhovid, V.M.~Lobashev, A.A.~Nozik, V.S.~Pantuev, V.I.~Parfenov, A.K.~Skasyrskaya, F.V.~Tkachov, S.V.~Zadorozhny}
\affiliation{Institute for Nuclear Research of Russian Academy of Sciences, Moscow, Russia}
\date{\today}

\begin{abstract}
An electron antineutrino mass has been measured in tritium $\beta$-decay in the Troitsk $\nu$-mass experiment. 
The setup consists of a windowless gaseous tritium source and an electrostatic electron spectrometer. 
The whole data set acquired from 1994 to 2004 was reanalyzed. A thorough selection of data 
with the reliable experimental conditions has been performed. We checked every known systematic 
effect and obtained the following 
experimental estimate for neutrino mass squared $m_{\nu }^{2}=-0.67\pm 2.53\;\mathit{eV}^{2}$. 
This gives an experimental upper sensitivity limit of $m_{\nu}<2.2\, \mathit{eV},\ 95\text{\%}\;\mathit{C.\,L.}$ 
and upper limit estimates ${m}_{\nu }<2.12\, \mathit{eV},\ 95\text{\%}\;\mathit{C. L.}$ for Bayesian statistics and ${m}_{\nu
}<2.05\, \mathit{eV},\ 95\text{\%}\;\mathit{C. L.}$ for the Feldman and Cousins approach. 

\end{abstract}

\maketitle

\section{Introduction} \label{sec:intro} 
The standard model of particle physics assumes zero mass for all neutrino 
flavors. However, the discovery of neutrino oscillations in experiments with solar, 
atmospheric and reactor neutrinos gives a strong evidence of a nonzero neutrino 
mass~\cite{oscill}. Oscillation parameters allow one to estimate the difference 
of mass squared values which give only the lower limit on neutrino eigenstate 
masses. The question of absolute values is still open. The most 
attractive methods to obtain an absolute mass value are neutrinoless double 
beta\d decay ($2\beta 0 \nu$) in even - even parity transitions in some 
nuclei (the probability of such a process depends on neutrino mass) and the method 
which measures the highest edge of electron energy spectrum in 
$\beta$ decay.    
In the former case, the decay is possible only if neutrinos are of the Majorana type, 
while in the latter case the experiment gives a model independent estimation 
of electron antineutrino mass irrespective of its type, Majorana or Dirac.

The measurement of the electron spectrum in tritium $\beta$ decay  
is one of the most precise direct measurements of neutrino mass. 
This type of measurements was utilized in the Troitsk and Mainz experiments. 
In 2003 the Troitsk group, having analyzed 
about half 
of the accumulated statistics, presented the upper limit for the neutrino mass at 95\% $m(\nu_e)<$\, 2.05 eV~\cite{troitsk2003}.
This result was obtained by excluding some additional small structure 
with unclear origin close 
to the spectrum end point.   
The  Mainz group in 2005 
published the final result of their search for neutrino mass~\cite{mainz}. They measured 
an upper limit of $m(\nu_e)\leqslant$\, 2.3 eV. 

In the paper we present a complete result of the Troitsk $\nu$-mass 
experiment. We reexamined the whole set of measurements  
reassessing the data quality and our 
knowlege of all the experimental conditions. Measurements with unstable or 
unclear conditions were removed. Some of the experimental corrections were 
reexamined. For each run of measurements we evaluated with the best known 
precision different experimental parameters, in particular, column density 
in a gaseous tritium source. In the current analysis we used a new method of 
quasioptimal weights~\cite{tkachev} to fit the measured electron spectrum. 
The obtained results were 
also compared with  the standard fitting procedure 
based on the {\scriptsize MINUIT} package~\cite{minuit}.  The two methods agree within statistical errors.

The paper is organized as follows:  
in Sec.~\ref{experiment} we briefly describe the experimental setup and 
measurement procedure. Analysis details are presented in Sec.~\ref{analysis}.
 In Sec.~\ref{errors} we describe systematic uncertainty. The final results are
presented in Sec.~\ref{results}, and  
 in Sec.~\ref{conclusions} we conclude.

\section{Experimental setup and procedure of measurements}
\label{experiment}
\subsection{Experimental Setup}
The choice of tritium as a $\beta$ decay  source is guided by its long half-life 
time (about 12.3 years), which garantees a long stability during the 
measurement time. Relatively low energy (the maximum electron energy is about 18.6 keV) makes 
it possible to use an electrostatic spectrometer. The simplicity of electron shells 
in molecular tritium allows one to calculate corrections on excited 
states in the molecules T$^3$He or H$^3$He, which are produced as  final states 
after the decay.
\begin{figure*}
\centering
\includegraphics[width=\textwidth]{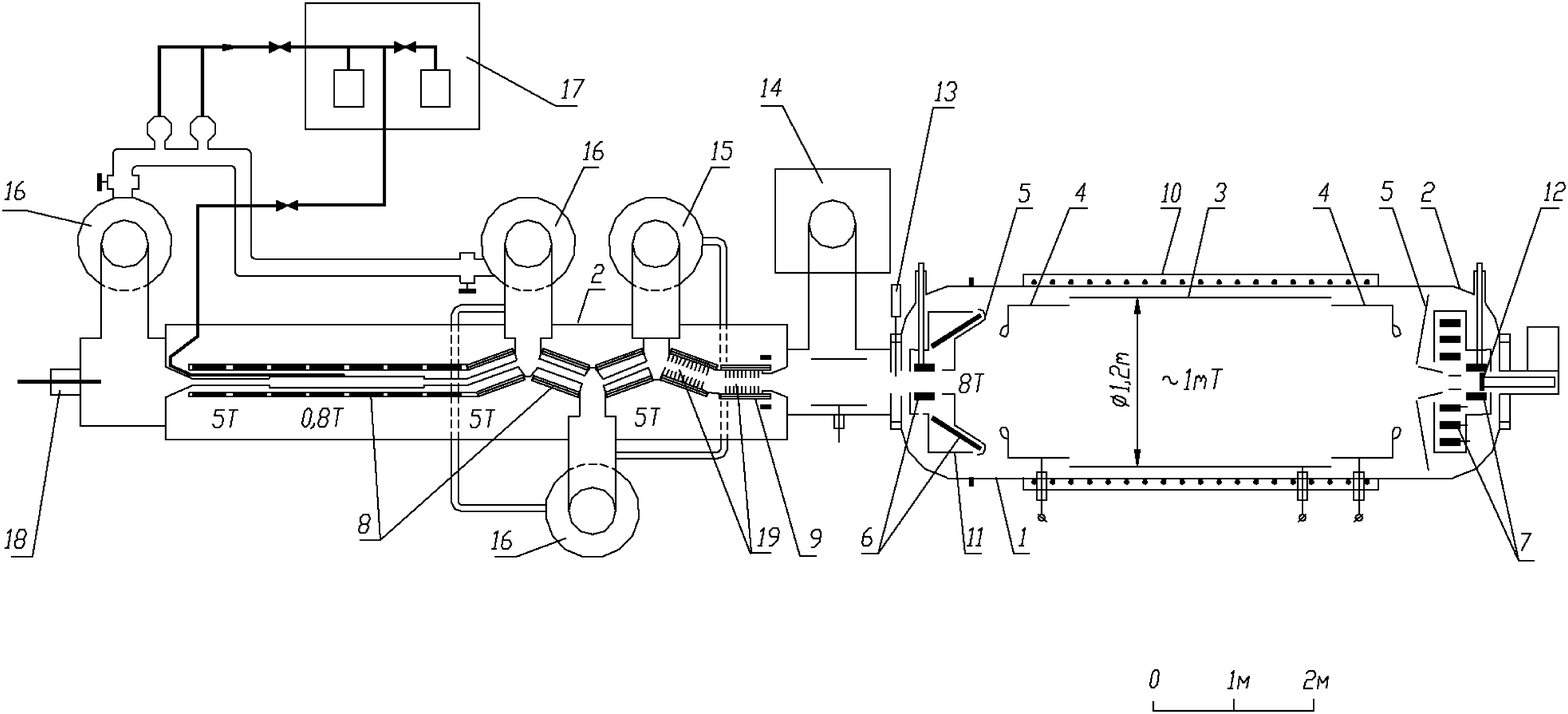}
\caption{Diagram of the installation:1,2 -- vacuum volume; 3,4 -- electrostatic system; 5 -- ground electrode; 6-9 -- superconducting coils; 10 -- warm solenoid; 11 -- Nitrogen shield; 12 -- Si(Li) detector; 13 -- emergency valve; 14 -- magneto-discharge pump; 15,16 -- mercury diffusion pumps; 17 -- tritium purification system; 18 -- electron gun; 19 -- argon trap.}
\label{setup}
\end{figure*}
The Troitsk experiment has two major features: the$\beta$ spectrum was measured 
by an integrating electrostatic spectrometer with adiabatic magnetic collimation, and  
a windowless gaseous tritium source (WGTS)~\cite{losalamos} was used as a volume for  $\beta$ decays.  The spectrometer allows one to get resolution of 3-4 eV, while the WGTS minimizes 
distortions of the electron spectrum. The setup is shown in Fig.~\ref{setup}, and details can be found in~\cite{troitsk2003}. 

Tritium gas is injected into a long pipe of WGTS in the axial magnetic field of about 0.8\,T, 
where tritium partially decays. At both ends of the pipe there are superconducting coils which form magnetic plugs with a field up to 5\,T. The reason for this magnetic field configuration was to avoid electron acceptance from tritium decays from the pipe wall.   
Electrons are transported via a zigzag-type transport system to the spectrometer, while residual gas and ions are 
pumped by a differential pumping system. After some purification the gas returns again to the pipe.

At the entrance of the spectrometer the magnetic field is formed by superconducting coils of $B_{max}=$8\,T; in the middle of the spectrometer the magnetic field drops to about 
$B_{min}\approx$1\,mT. Magnetic field lines are collected again by a 3\,T magnet with a Si(Li) 
counter inside. This configuration of the magnetic field collimates electrons in  
such a way that a transverse component of the electron momentum becomes small near the middle 
of the spectrometer (analyzing plane) and 
the electron angular distribution along the spectrometer axis is limited by 
a small  value of $\delta=B_{min}/B_{max}$. In the analyzing plane there is also a 
strong electrostatic retarding  field oriented against the electron direction. 
Only the electrons with energy above the retarding field will pass the barrier, while all the 
other electrons with smaller energy will be reflected. By changing the electrostatic potential we can scan and get an integrated electron spectrum.

Electrons at the far end of the spectrometer are counted by an Si(Li) detector with a sensitive area 
 of about 17 mm in diameter . The signal amplitude and its arrival time are digitized and readout by a computer and online KAMAK electronics with a fixed dead time of 7.2 $\mu$sec.  
\subsection{Procedure}
The measurement procedure was as follows: the integrated yield of $\beta$-electrons near 
the end point of the spectrum was scanned by changing the electrostatic potential in the spectrometer to a  
range between 18000 and 18900 volts. There were 60-80 set points with a measurement time 
of 10 to 200s depending on the 
count rate at the set point. 
The sequence of points in potential values was forward and reverse and random as well.  
To control the intensity in the WGTS, every 15min there was a  monitor point 
measurement at 18000\,V, where the counting rate is large. 

The data format was as follows: at the beginning of each scan we checked the readiness 
of the electronics and the high voltage system. Then we started the scan by varying the electrostatic potential.
 For each set point high voltage was checked to be within 0.2\,V of the required value.
 The value of this deviation was checked every second and recorded in the file for further 
offline corrections. At the end of each scan we wrote the pressure in the WGTS and started the next set 
in the opposite direction. 

During the measurements we controlled and recorded the temperatures of cooling helium and 
superconducting magnets. About every 2h we measured hydrogen isotope concentrations in the WGTS. 
\section{Data analysis}
\label{analysis}
\subsection{Data selection and experimental corrections}
During the preliminary data selection and analysis we checked the consistency of the mean 
count rate at each set point. Analysis shows that there are increases of the counting rate. This effect is induced by a local discharge from 
tritium decays inside the spectrometer (there is a small but finite probability 
for molecules from the WGTS to penetrate to the spectrometer) or from electrons which escape 
from magnetic ``traps'' inside the spectrometer. A special algorithm was developed 
to find these bunches and exclude that time interval from the analysis.
 After that, we checked the distribution of time intervals between events, which followed 
the Poisson distribution and looked like a pure exponentially falling distribution. At 
set points where the intensity was large and it was hard to distinguish such bunches, we 
extrapolated from points with a low counting rate. 

Data were corrected for signal pileup and for electronics dead time. In the final analysis 
we used points only above 18400\,V, where these corrections are small, except for the monitor 
points at 18000\,V.

Files with a full set of measurements were then checked for stability of the counting rate 
at the monitor point within 10\% from the average value. This allows one to control 
the stability of isotope contents in the WGTS, avoiding a sudden change caused by 
the purification system. Points with large high voltage offsets were also removed. 
Special care was taken to keep only the runs where precise measurements of the column density 
in the WGTS were performed (see below).
\subsection{Method of quasioptimal weights}
The fit of parameters in the previous analysis~\cite{troitsk2003} was done by means of 
the standard {\scriptsize MINUIT} package which uses the
method of least squares. Yet the effectiveness of such a method of parameter estimation 
is not guaranteed at a low number of counts
 where the distribution is Poissonic rather than Gaussian. To account 
for that problem a method of
quasioptimal weights in Ref.~\cite{tkachev} was implemented. This quite general procedure uses 
a wellknown method of moments as a basis.
The method of moments is simple, reliable, and  analytically 
transparent, but its effectiveness can be low. A way to eliminate the latter drawback was described in the same article~\cite{tkachev}. The general scheme of the method is as follows.

First one has to choose weight functions  $\phi _{i}(X_{i})$ of measured 
values  $X_{i}$ (in our case
$X_{i}$ are count numbers for different retarding potentials on the electrode). 
Then one should calculate the weighted
average for the data set and the corresponding average over the fitting curve, which depends 
on the parameters $\theta$ being
estimated:
\begin{center}
\begin{gather}
\langle \phi \rangle _{\exp }=\frac{1}{N}\sum _{i=1}^{N}\phi _{i}(X_{i}), \nonumber \\ 
\langle \phi \rangle_{\mathit{th}}=\frac{1}{N}\sum _{i=1}^{N}\langle \phi _{i}(\theta)\rangle _{\mathit{th}},
\label{eq:beta}
\end{gather}
\end{center}
where $N$ is the number of points in the file.

Requiring  $\langle \phi \rangle _{\exp }=\langle \phi \rangle _{\mathit{th}}$, we get equations on  
$\theta $. If
one gets a number of different weights  $\phi $ (equal to the number of parameters  $\theta $), 
it is possible to get
the system of equations, whose solution  is the estimate of $\theta $. Variation of  
$\langle \phi \rangle
_{\mathit{th}}$ gives error estimation for parameters. As for the choice of weights, there is a simple explicit 
formula for optimal weight, which gives minimal variation of  estimation of parameters 
(Rao-Cramer bound). This
formula involves the unknown values of $\theta $, but deviation of 
variance from the Rao-Cramer 
minimum is quadratic with respect to the deviation of weights from the optimal expression; thus it makes practical sense to use not the exact optimal weights
based on unknown ``real'' values of parameters, but quasioptimal ones based on parameter values that are 
close to the ``real'' ones. A poor choice of the weight would not affect 
the consistency of the method, but the variance would be larger than for the optimal weight; in other words, the resulting estimate would be suboptimal but still correct.

Efficiency of the method and stability of its program implementation (a robust code written in statically type-safe component
pascal) were tested by comparison with the most commonly used methods. 
Statistical tests showed that the
efficiency of the method of quasioptimal weights is equal to that of the method of maximum 
likelihood (which is known to
give the best effectiveness in such cases). Direct comparison of the parameter  
obtained using {\scriptsize MINUIT}
(the {\scriptsize JMINUIT} package was used~\cite{jminuit}) to the quasioptimal weights method showed no discrepancies, within the calculation uncertainties.
\subsection{Spectrum and corrections}
In our experiment we measured an integrated electron spectrum. Thus, we have to start with 
an unmodified theoretical $\beta$ spectrum of tritium decay and integrate it with experimental 
resolution. The spectrum is also distorted by electron interactions in the WGTS. There is also another  
set of corrections, and as a result, the analysis has several steps.

The electron energy spectrum in $\beta$ decay is described by the following wellknown expression:
\begin{center}
\begin{eqnarray}
S(E, E_0, m_{\nu})=N\cdot F(Z,E)\cdot(E+m_e)\cdot p\cdot (E_0-E)^2 \nonumber \\ \times \sqrt{1-\frac{m^2_\nu}{(E_0-E)^2}} ~
\label{eq:beta}
\end{eqnarray}
\end{center}
where $E$, $p$, and $m_e$ are the electron kinetic energy, momentum, and mass; $m_{\nu}$ is the neutrino mass; 
$E_0$ is the spectrum energy edge in the case $m_{\nu}=0$; $N$ is the normalization constant and 
$F(Z,E)$ is the so-called Fermi function, which induces an electrostatic correction to 
the charge $Z$ of the residual nucleus~\cite{fermi}. 

Equation~\ref{eq:beta} depends on $m^2_{\nu}$, and 
 the preliminary analysis of both experiments, Troitsk and Mainz, has shown that 
experimental estimations on $m^2_{\nu}$ may get negative values. Besides statistical 
fluctuations, such behavior  could be attributed to some experimental systematics with 
unknown origin which moves the spectrum end point beyond its maximum value $E_0$. To account 
for this effect, Eq.~\ref{eq:beta} should be extended to negative ranges 
of $m^2_{\nu}$.  
We also checked different methods of such an extension and found a weak dependence of the result on the actual choice.  
Finally, we chose the method used in the Mainz experiment~\cite{mainz}. 

\begin{figure}[htb]
\centering
\includegraphics[width=0.55 \textwidth]{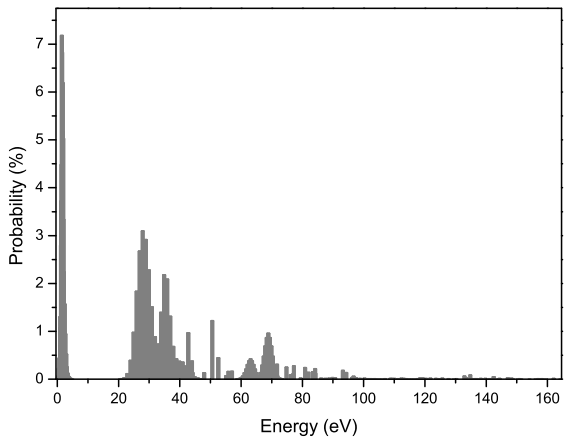}
\caption{Final states spectrum in molecule T$^3$He~\cite{states}. Every bin is a summand in eq.~\ref{eq:states}.}
\label{fig:states}
\end{figure}

Often after the decay of a tritium nucleus, the final molecule T$^3$He will not go to 
its ground state; thus, we have to sum over all final states $i$, and Eq.~\ref{eq:beta} 
should be replaced by the sum
\begin{equation}
S(E)=\sum_i N(E, E_0-E_i)\cdot P_i , 
\label{eq:states}
\end{equation}
where $E_i$ is energy of the excited state and $P_i$ is its probability. The summation is done 
over the excited spectrum divided into a set of narrow bins, as shown in Fig.~\ref{fig:states}, keeping 
the sum of  $P_i$ equal to 1.  
Unfortunately, this spectrum was not measured experimentally with good accuracy; thus, we 
have to use theoretical model calculations. We use a generated spectrum from 
Ref.~\cite{states}. For comparison, we also checked a few other models and found that a 
final result on the square value of the neutrino mass does not change much and stays within  
our estimation of the total systematic uncertainty. 

\begin{figure}[htb]
\centering
\includegraphics[width=0.55 \textwidth]{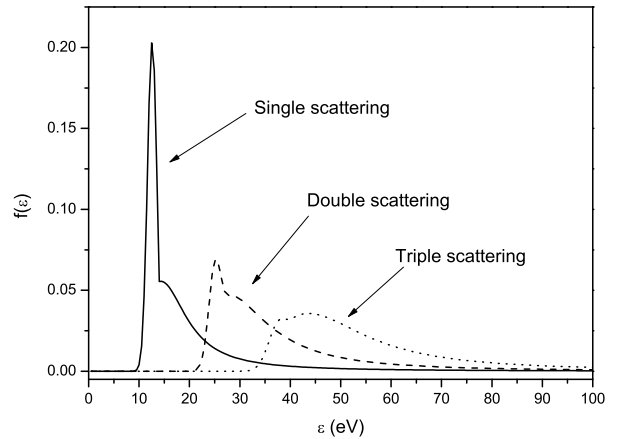}
\caption{The shape of the electron energy loss spectrum in tritium~\cite{resolution}. Different curves correspond to electron single, double, and triple scattering.}
\label{fig:eloss}
\end{figure}

Electrons in the WGTS suffer from scattering on tritium molecules. To account for  
such an effect, we have to convolute Eq.~\ref{eq:states} with the energy loss  
function. We use a detailed analysis of this function, which was performed 
in~\cite{resolution}. In Fig.~\ref{fig:eloss} we show the electron energy loss 
spectrum in tritium for single, double, and triple scattering. The results for double and 
triple scattering were calculated as a convolution of a single loss spectrum with itself.

Multiple scattering should follow the Poisson distribution; thus, we can write the 
probability for scattering of the order of $k$ as
\begin{equation}
P_k=\frac{X^ke^{-X}}{k!}. 
\label{eq:scatt}
\end{equation}
Here $X=\int \limits_0^L\frac{dl}{\lambda(l)}=\int \limits_0^L\sigma_{tot}n(l)dl$ 
is the ratio of the electron path length in the gas to a mean free path, where $L$ is the pass length, $n(l)$ is the gas density at point $l$, and $\sigma_{tot}$ is the total inelastic cross section. In practice, in our calculations we considered  
up to the triple scattering processes only: at a typical value of $X=$ 0.35,  $P_4$ = 0.00044. 

Electrons produced in different parts of the WGTS have different $X$,  and so we  
average each value of $P_k$ over the path length . This averaging may be performed  
analytically. Suppose all electrons move exactly along the magnetic field 
lines which in the WGTS are directed along its length. In a volume  element 
of the pipe, $Sdl$, the number of molecules is $dN=Sn(l)dl$, where $S$ is the 
pipe cross section, $n$\,\textendash \,the gas density, and  $l$\,\textendash \,the coordinate 
along the pipe. If $dX=\sigma_{tot}n(l)dl$, 
then $dN=S\diagup \sigma_{tot}n(l)dX=C \cdot dX$, where $C$ is a constant. 
The average probability for a path with no scattering will be
\begin{equation}
<P_0>=\frac{C\int_o^{N_{tot}}e^{-X}dN}{C\int_o^{N_{tot}}dN}=\frac{\int_o^{X_0}e^{-X}dX}{\int_o^{X_0}dX}=\frac{1}{X_0}(1-e^{-X_0}). 
\label{eq:p0}
\end{equation}
Here $X_0$ is the total length expressed in units of the mean free path. In a similar 
way we get
\begin{gather}
<P_1>=\frac{1}{X_0}(1-e^{-X_0})-e^{-X_0}, \nonumber \\ 
<P_2>=\frac{1}{2X_0}(2-e^{-X_0}(X_0^2+2X_0+2)),\nonumber \\
<P_3>=\frac{1}{6X_0}(6-e^{-X_0}(X_0^3+3X_0^2+6X_0+6)).\, \,\,  
\label{eq:average}
\end{gather}

In addition, we have to take into account electron circular motion which increases electron path length while they are 
moving in the magnetic field. This increase 
 depends on the orientation of the electron momentum vector relative to the magnetic 
field direction.  The magnetic field does not change the absolute value of 
the electron velocity $V$, but changes its direction, keeping the velocity 
longitudinal component $V_Z$ constant. Thus, we can write the expression for 
time  which is needed for the electron to cover a distance $z$ along the pipe as $t=z/V_z$. 
The total electron path is $D=V\cdot t$, and we can write
\begin{equation}
\frac{X}{X_0}=\frac{D}{z}=\frac{V}{V_z}=\frac{1}{\cos\theta}, 
\label{eq:tangle}
\end{equation}
where $\cos \theta$ is the angle between the electron velocity and the magnetic field direction. 

We calculated $P_i$ taking into 
account the fact that only a fraction of electrons will pass to the spectrometer.
The results for corrections on electron 
magnetic winding were approximated by linear functions:
\begin{gather}
P_0=<P_0>\cdot(0.9996-0.0398\cdot X_0), \nonumber \\ 
P_1=<P_1>\cdot(1.0854-0.0460\cdot X_0), \nonumber \\ 
P_2=<P_2>\cdot(1.1595-0.0567\cdot X_0),\nonumber \\
P_3=<P_3>\cdot(1.2398-0.0682\cdot X_0).
\label{eq:pfinal}
\end{gather}
Strictly speaking, the linear approximation is our arbitrary choice, but as we have  
found, the contribution of higher orders is less than 0.05\% for our range of the parameter $X_0$.

\begin{figure}[htb]
\centering
\includegraphics[width=0.55 \textwidth]{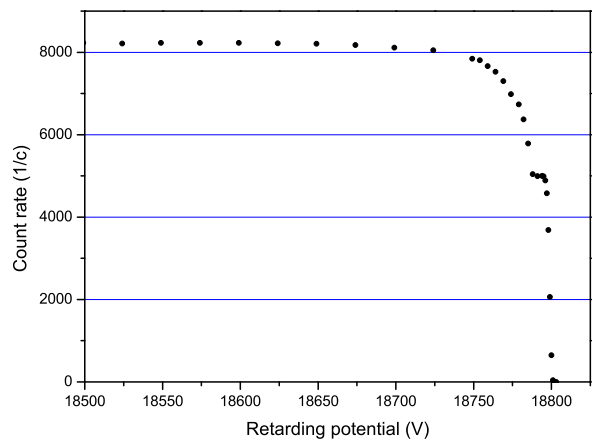}
\caption{Integrated spectrum from the electron gun versus the applied potential on the spectrometer electrode. The WGTS pipe is filled with tritium. The voltage on the electron gun cathode is around 18800\,V.}
\label{fig:step}
\end{figure}

Direct measurement of gas density in the WGTS pipe with the required precision is impossible. During data taking, for each file we measure the intensity in the monitoring point at the spectrometer potential $U$ = 18000\,V. At such a voltage, a significant portion of the electron spectrum will pass the spectrometer with a relatively large counting rate $N_{mon}$. This rate is proportional to the total amount of tritium in the source pipe. An additional mass 
analyzer directly connected to the WGTS, at the same time measures partial concentrations of T$_2$, TH, and H$_2$ molecules. From this measurement we calculate the ratio $P_T$ of tritium atoms to the total number of hydrogen isotope atoms. Introducing an additional calibration constant $A$, we can write the relation
\begin{equation}
X_0=A\cdot \frac{N_{mon}}{P_T}. 
\label{eq:x0}
\end{equation}
The calibration constant $A$ depends on many experimental conditions, such as magnetic field configuration or temperature in the pipe, but during a particular run, it remains constant within the systematic uncertainties. We find the value of $A$ experimentally using an electron gun mounted at the rear end of the WGTS. The gun produces a monochromatic beam of electrons in the energy range of up to 20\,keV which pass through the whole WGTS pipe. With no gas in the pipe the gun allows us to measure the transition function or resolution of the spectrometer. When the pipe is filled with gas the integrated spectrum from the gun in the spectrometer changes, Fig.~\ref{fig:step}.    
There is a sharp edge of the spectrum to the right located at the gun potential. To the left of this edge, first we see a flat step with a width of about 12\,V, which corresponds to electrons with no scattering in the pipe. Then, there is  another rise from electrons which lost at least 12\,eV after a single scattering, compare this behavior with the energy loss spectrum in Fig.~\ref{fig:eloss}. The integral of the spectrum at spectrometer potential values below 200\, from the right edge corresponds to all electrons with or without scattering. The ratio of the magnitude of the right step to the total number of electrons in the left part of the spectrum is equal to the probability with no scattering, which is related to $X_0$. Taking into account the correction for track winding for electrons from the gun we can solve Eq.~\ref{eq:x0} for parameter $A$. Such calibration measurements for parameter $A$ were performed for each run. The runs which did not have these calibrations were rejected.

At an early stage of the experiment we found that there is an additional contribution to the spectrum from electrons which are trapped in the WGTS. More than 90\%  of $\beta$ electrons which were produced at a large angle relative to the axial magnetic field cannot escape because of strong fields which work as magnetic plugs at both ends of the WGTS .  In the adiabatic regime the maximum electron angle relative to the magnetic field for the electrons to escape through the plug can be found from
\begin{equation}
\sin\,\alpha_{max}=\sqrt{\frac{B_S}{B_T}}=\sqrt{\frac{0.8}{5}}=0.4,
\label{eq:plug}
\end{equation}
where $B_S$ is the field in the pipe and $B_T$ is the field in the transport system. Thus we get $\alpha_{max}$=23.5$^{\circ}$.

\begin{figure}[htb]
\centering
\includegraphics[width=0.55 \textwidth]{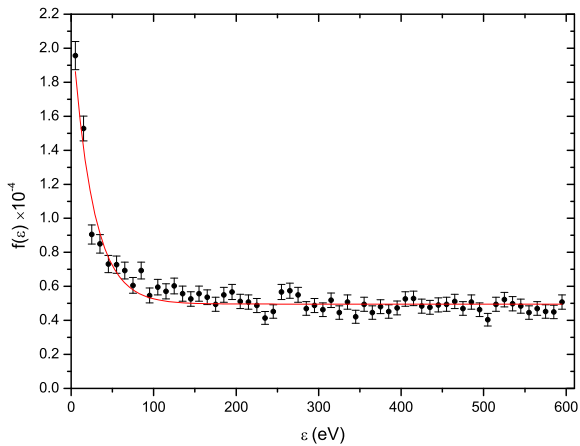}
\caption{Energy loss spectrum, $\varepsilon=E_{in}-E_{fin}$, for electrons which were trapped in WGTS but after scattering reached the spectrometer. The bin size is 10\,eV. The solid line is the analytic approximation of the losses.}
\label{fig:trapped}
\end{figure}

Trapped electrons suffer from multiple reflections from the magnetic plugs. In general, they cannot escape the trap. However, electrons may scatter on molecules in the WGTS, change their angle relative to the magnetic field, and be transported to the spectrometer. Such an effect is electron diffusion in the surrounding gas to the transport system phase space. The portion of these electrons which reaches the spectrometer is only about $10^{-4}$ from the electrons within the acceptance. Nevertheless, we have to account for this effect because the energy loss spectrum for trapped electrons is very different. We did Monte Carlo simulation for tritium decays in the WGTS. The total number of the generated electrons was $10^7$, the number of electrons which finally got to the spectrometer was 9800. In Fig.~\ref{fig:trapped} we show the energy loss spectrum for these electrons.      
Simulated results were approximated by an analytic function 
\begin{equation}
trap(\varepsilon)=1.86\cdot 10^{-4}\cdot exp(-\frac{\varepsilon}{25})+5.5\cdot 10^{-5}
\label{eq:analyt}
\end{equation} 
shown by a solid line in Fig.~\ref{fig:trapped}. The final energy loss function could be written as
\begin{equation}
Tr(\varepsilon)=P_0\delta (\varepsilon)+P_1f_1(\varepsilon)+P_2f_2(\varepsilon)+P_3f_3(\varepsilon)+trap(\varepsilon) ,
\label{eq:trapped}
\end{equation}
where $P_i$ are mean probabilities to scatter $i$ times from Eq.~\ref{eq:pfinal} and $f_i(\varepsilon)$ are energy loss distribution functions for $i$th scattering. 

The electron spectrum should be integrated with a resolution function which is defined by the following equation~\cite{resolution}: 
\begin{center}
\begin{equation}
R(U,E)=
\begin{cases}
0& E-U<0, \\
\frac{1-\sqrt{1-\frac{E-U}{E}\frac{B_S}{B_A}}}{1-\sqrt{1-\frac{\Delta E}{E}\frac{B_S}{B_A}}} & 0\leq E-U\leq \Delta E,\\
1 & E-U \geq \Delta E,
\end{cases}
\label{eq:resolution}
\end{equation}
\end{center}
where $E$ is the electron energy, $U$ is the spectrometer electrode potential, $\Delta E=\frac{B_A}{B_0}E$, $B_A$ is the magnetic field in the spectrometer analyzing plane, $B_S$ is the magnetic field in the WGTS pipe, and $B_0$ is the field in the pinch magnet at the entrance of the spectrometer. 

\begin{figure}[htb]
\centering
\includegraphics[width=0.55 \textwidth]{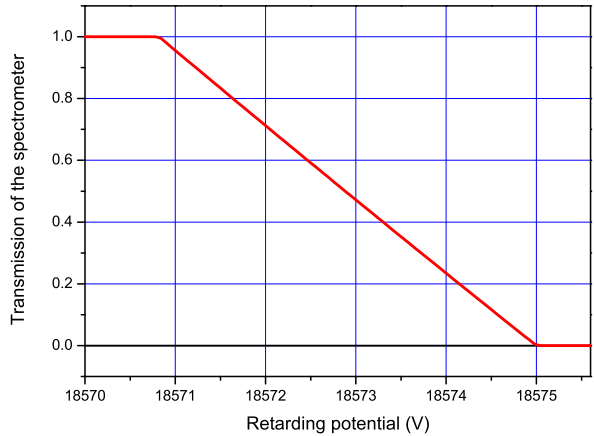}
\caption{Resolution function for electrons with energy of 18575 eV. The curve corresponds to the magnetic field ratio $B_A / B_0$=2.26\,10$^{-4}$.}
\label{fig:resolution}
\end{figure}

We use this analytic form for the resolution function which depends only on field configurations. To justify the validity of Eq.13, we performed a full 
simulation with the nominal electrostatic and magnetic 
fields in the realistic geometry. We found that an analytic 
representation of the transmission function by 
Eq.13 describes very well results of the simulation.
The experimental resolution, or transmission function, was also 
measured with the electron gun and the results agree with the theoretical estimate with errors which are determined based on the stability of the high voltage system. These errors are treated as systematic uncertainties. The resolution function is shown in Fig.~\ref{fig:resolution} and looks almost linear. 

Finally, we get the following expression for the experimental integrated electron spectrum:
\begin{equation}
Sp(U)=N\cdot \int \big[S(E,E_0,m_{\nu}^2)\otimes Tr(E) \big]\cdot R(U,E)dE + bkgr ,
\label{eq:spectrum}
\end{equation}
where $S(E,E_0,m_{\nu}^2)$ is the electron spectrum from $\beta$ decay [Eq.~\ref{eq:states}], $Tr(E)$ is the energy loss spectrum [Eq.~\ref{eq:trapped}], $R(U,E)$ is the resolution function [Eq.~\ref{eq:resolution}], and bkgr is experimental background. 

In the data analysis we use four free parameters: $m_{\nu}^2$, $E_0$\,\textendash \,the spectrum energy edge for the case $m_{\nu}=0$, $N$\,\textendash \,the normalization constant and bkgr.   

\subsection{Summing-up files}
Each data file was measured within about 2 h. During one run of measurements an effective column density in the WGTS may vary by 10\% from file to file. To add files with different density we use the following procedure (for simplicity we present an example for two files):
\begin{eqnarray}
Sp_1(U)+Sp_2(U)= 
N_1\int S(E)\otimes Tr_1\cdot R(U,E)dE\nonumber \\ 
+N_2\int S(E)\otimes Tr_2\cdot R(U,E)dE\nonumber \\  
=\int S(E)\otimes (N_1Tr_1+N_2Tr_2)\cdot R(U,E)dE\nonumber \\
=2\int S(E)\otimes (\frac{N_1\cdot P_0^1++N_2\cdot P_0^2}{2}\delta(\varepsilon) \nonumber \\ 
+\frac{N_1\cdot P_0^2+N_2\cdot P_0^2}{2}f(\varepsilon)+\dotsm )\cdot R(U,E)dE, \,\,\,\,\,\, 
\label{eq:addfiles}
\end{eqnarray}
where $N_1$ and $N_2$ are normalization constants for each file from a fit by Eq.~\ref{eq:spectrum}. In this procedure, over many files, we actually average probabilities for multiple scattering:
\begin{equation}
P_i=\frac{\sum_{j=1}^n P_i^j\cdot N^j}{\sum_{j=1}^n N^j} \,\,\, (i=0-3).
\label{eq:averP}
\end{equation}
\section{Systematic errors}
\label{errors}
The main source of
systematic uncertainties is the uncertainty in the estimation of the WGTS column thickness
$X_0$. During one run the value of the source thickness is constantly varying (in the bounds of
5\%-10\% from the mean value).  
For each data file with a duration of 2-2.5 h, $X_0$ was measured using Eq.~\ref{eq:x0}.  The error of the count rate in the monitor point is negligibly small (less than 0.1\%). The error for the tritium concentration  mainly comes
from its drift during the file (about 1.5\%). The error for coefficient $A$ is calculated from its
estimation procedure and is 1.5\%. Therefore, we use 3\% as a conservative error of $X_0$. 

The second contribution to the systematic uncertainty comes from the final state spectrum of T$^3$He, Fig.~\ref{fig:states}. As mentioned above, there is no direct experimental measurements of this spectrum, and we have to use theoretical estimates.
The influence of the uncertainty in the final state spectrum was investigated in ~\cite{troitsk2003} and was found to be 0.7 eV$^2$  in the neutrino mass squared determination.

The error in the trapping-effect estimation arises from the uncertainty in the cross sections of the electron interaction 
with a tritium  molecule. This error was taken as 20\% of the full amplitude of the trapping effect. The influence of 
this error on the neutrino mass squared is calculated individually for each run and varies within 0.3-0.5~\,eV$^2$.

An additional uncertainty comes from the instability of the potential on the main spectrometer electrode, which is
less than 0.2 eV. The shift of the squared neutrino mass due to such an effect was estimated in~\cite{losalamos}. According to the 
formula derived in this work, $\Delta m_{\nu }^{2}=-2\sigma _{E}^{2}$. Thus, the shift of the
neutrino mass squared is less than 0.08 eV$^2$. It
should be noted that the efficiency of the Si detector and the absolute value of the retarding potential on the spectrometer electrode do not affect the estimate
of the mass because the normalization factor and the end point energy are free parameters.

 Estimates of statistical and systematic errors were made for each run. To estimate the effect 
of the source thickness uncertainty, the following procedure was used for each run:
\begin{enumerate}
\item We fit the spectrum with an average value of the source thickness and estimate  the squared neutrino mass  $\langle
m_{\nu }^{2}\rangle $.
\item Then we fit with the thickness value shifted by its error ($\pm$ 3\%)  to get the new estimates for $\langle m_{\nu }^{2}\rangle
_{\mathit{\pm shift}}$.
\item The averaged difference  $|\langle m_{\nu }^{2}\rangle -\langle m_{\nu }^{2}\rangle _{\mathit{\pm shift}}|$ is taken as a
systematic uncertainty from the source thickness.
\item The systematic uncertainty from the trapping is calculated in a similar way.
\end{enumerate}

A small error also comes from the processing of preliminary data. There, detector dead time and overlapping events
 are taken into account. Corrections for dead time and
overlapping become visible only at relatively high count rates when the retarding
potential is lower than 18400 V. These points were not used in the analysis of the spectrum.

The sources of systematic uncertainty are:
\begin{enumerate}
\item Uncertainty of source thickness.
\item Final state spectrum ambiguity.
\item Uncertainty in parameters of the trapping effect.
\item Instability of the retarding potential.
\end{enumerate}
All errors are summed quadratically.

\section{Results}
\label{results}

\begin{table*}
\centering
\begin{tabular}{|m{3cm}|m{2.cm}|m{1.6cm}|m{1.6cm}|m{1.6cm}|m{1.6cm}|m{1.6cm}|m{1.6cm}|m{1.6cm}|}
\hline
\centering{\bfseries Run} &
\centering{\bfseries Date}\\
{month.year} &
\centering {\textbf{$m_{\nu}^2$}} &
\centering {\textbf{$\sigma_{stat}$}} &
\centering {\textbf{$\sigma_ X$}} &
\centering {\textbf{$\sigma_{trap}$}} &
\centering {\textbf{$\sigma_{syst}$}} &
\centering {\textbf{$\chi^2/d.o.f.$}} &
{\textbf{$\chi^2_S/d.o.f.$}} \\
\hline
{22}  &
\centering 06.1997 &
\centering -7.55 &
\centering 9.89 &
\centering 1.1 &
\centering 0.34 &
\centering 1.34 &
\centering 0.796 &
0.814\\
\hline
{ 23} &
\centering 12.1997 &
\centering 2.53 &
\centering 4.57 &
\centering 1.31 &
\centering 0.352 &
\centering 1.52 &
\centering 1.043 &
1.07 \\
\hline
{24},{ first part} &
\centering 01.1998 &
\centering -1.31 &
\centering 4.32 &
\centering 1.35 &
\centering 0.318 &
\centering  1.55 &
\centering  0.923&
0.964\\
\hline
24, second part &
\centering 02.1998 &
\centering -5.44 &
\centering 4.98 &
\centering 1.48 &
\centering 0.342 &
\centering 1.67 &
\centering 1.026 &
1.041\\
\hline
25 &
\centering 06.1998 &
\centering -0.11 &
\centering 7.35 &
\centering 1.57 &
\centering 0.378 &
\centering 1.76 &
\centering 0.847&
0.739 \\
\hline
28 &
\centering 05.1999 &
\centering 2.60 &
\centering 6.99 &
\centering 1.82 &
\centering 0.4 &
\centering 1.99 &
\centering 1.421 &
1.496 \\
\hline
29 &
\centering 10.1999 &
\centering -0.51 &
\centering 7.50 &
\centering 1.94 &
\centering 0.416 &
\centering 2.10 &
\centering 1.268 &
1.456\\
\hline
30 &
\centering 12.1999 &
\centering 3.14 &
\centering 8.31 &
\centering 2.04 &
\centering 0.434 &
\centering  2.19 &
\centering  1.523&
1.327 \\
\hline
31 &
\centering 12.2000 &
\centering -8.06 &
\centering 6.99 &
\centering 1.45 &
\centering 0.38 &
\centering  1.65 &
\centering  0.902&
0.943\\
\hline
33 &
\centering 06.2001 &
\centering 7.21 &
\centering 8.82 &
\centering 1.47 &
\centering 0.504 &
\centering 1.70 &
\centering 1.378 &
 \\
\hline
36 &
\centering 04.2002 &
\centering 1.91 &
\centering 6.72 &
\centering 1.37 &
\centering 0.322 &
\centering 1.57 &
\centering  1.356&
 1.379\\
\hline
\end{tabular}
\hfill{}
\caption{Results for the neutrino squared mass estimate for different runs. All values are in eV$^2$. 
Total systematic uncertainties are shown in the seventh column. The next to last column represents $\chi^2/d.o.f.$ obtained in each fit. The last  column, $\chi^2_S/d.o.f.$, demonstrates how much the $\chi^2$ value changes by introducing an additional steplike structure as was done in the previous analysis~\cite{troitsk2003}.}
\label{tab:data}
\end{table*}
Results of the analysis are presented in Table~\ref{tab:data}. Runs that were too short and runs where external parameters 
(mainly source
thickness) could not be estimated with the required precision were not used in the analysis. In particular, run 21 (May
1997) and all the earlier runs were not included because there were no calibrations done with an electron gun, and
consequently, the thickness value $X_0$ was unreliable. 

The final result and systematic uncertainty were obtained by  averaging over all runs weighted using statistical errors.  
Thus,
systematic uncertainty for individual run does not affect the overall estimate of the neutrino mass squared. As a result we get
\begin{equation*}
 m_{\nu }^{2}=-0.67\pm 1.89_{\mathit{stat}}\pm 1.68_{\mathit{syst}}\;\mathit{eV}^{2}.
\end{equation*}

After summing  errors in quadrature our estimate is
\begin{equation*}
m_{\nu }^{2}=-0.67\pm 2.53\;\mathit{eV}^{2}.
\end{equation*}

The result for the neutrino mass squared is negative but close to zero, within one sigma. For comparison, the result obtained earlier by our group~\cite{troitsk2003} is  $m_{\nu
}^{2}=-2.3\pm 2.5_{\mathit{stat}}\pm 2.0_{\mathit{syst}}\;\mathit{eV}^{2}$, or  $m_{\nu }^{2}=-2.3\pm
3.2\;\mathit{eV}^{2}$. An improved precision of the current analysis comes from the usage of four free parameters in the fit
(instead of six, as was done earlier with two additional parameters for a steplike structure) and an increase of the data amount. 
To decrease systematic uncertainties we also increase the low energy cut off of the data range from 18300\,V to 18400\,V.

Since the final $m_{\nu }^{2}$ value is slightly negative, one can derive an upper physical bound for the neutrino mass. There is no single universal way to do this. Many experiments published the Bayesian limits which were calculated from the measured $m_{\nu }^{2}$ value. It seems that for a value which is out of the physical region the most correct way would be to calculate the so-called sensitivity limit~\cite{tkachev2}. It uses error information but not the estimate itself; i.e., it is not sensitive to how negative the estimate is. 
In our case, this limit is calculated in the following way:
\begin{gather*}
m_{\nu }^{2}<2.53\times 1.96=4.96~ \text{eV}^2,\; 95\text{\%}\;\mathit{C.\, L.},
\end{gather*}
where 1.96 is a standard multiplier for the 95\% confidence level. For the neutrino mass this gives  $m_{\nu
}<2.2\, \mathit{eV}$. The corresponding value obtained by the Mainz group was $m_{\nu }<2.4\,\mathit{eV}$~\cite{mainz}. 

The unified approach of Feldman and Cousins~\cite{feldman} and Bayesian methods yield the following upper limits for $m_{\nu }$:
\begin{gather*}
{m}_{\nu }<2.12\, \mathit{eV},\ 95\text{\%}\;\mathit{C.\, L.}\ \text{Bayesian},\\
{m}_{\nu}<2.05\, \mathit{eV},\ 95\text{\%}\;\mathit{C.\, L.}\ \text{Feldman and Cousins.}
\end{gather*}

The coincidence of our neutrino mass upper limit in the Feldman and Cousins approach with the result presented in 2003~\cite{troitsk2003} is accidental. In the current analysis the final error is smaller, but $m_{\nu }^{2}=-0.67\pm 2.53\;\mathit{eV}^{2}$ is less negative compared to the value of $m_{\nu }^{2}=-2.3\pm 3.2\;\mathit{eV}^{2}$ in~\cite{troitsk2003}.

We also want to stress that in our analysis there is no need for any additional structure, like a step, at the upper end of the $\beta$-electron spectrum, which made an unambiguous interpretation difficult. To confirm this, we performed additional fits with two extra parameters in an attempt to reproduce a steplike structure, as was done in the old analysis~\cite{troitsk2003}. In the last column of Table~\ref{tab:data} we present  $\chi^2_S/d.o.f.$ values for these fits. For run 33 the fit with the step did not converge
despite all our attempts. There is no significant change in $\chi^2$ values with such a step. Then, considering that the major difference between the two analyses is an estimate of  the source thickness, for two runs we manually decreased the source thickness by a few percent to the value used in the old analysis. After that, the step reappears at about the same position from the spectrum's upper end  and with a similar amplitude. Thus, we conclude that the reasons why the results of the new analysis differ from the old one are a more thorough  and careful file selection and a more complete 
accout of the experimental conditions.  

\section{Conclusion}\label{conclusions}
Data analysis was performed over a set of data taken from 1994 to 2004 by the 
Troitsk $\nu$-mass experiment. Very early runs and a few runs taken after 1997 were rejected due to the lack of full information on experimental conditions or missing calibrations. The knowledge of the total column density in the windowless gaseous tritium source appeared to be the most critical. Some statistics were added from the runs which were excluded from the previous analysis. 

For the analysis a new method of quasioptimal moments was used with Poisson statistics  of experimental points with a low counting rate. An experimental estimate for the neutrino mass squared is $m_{\nu }^{2}=-0.67\pm 2.53\;\mathit{eV}^{2}$. From this we obtain an upper sensitivity limit $m_{\nu}<2.2\, \mathit{eV},\ 95\text{\%}\;\mathit{C.\,L.}$ and upper limit estimates ${m}_{\nu }<2.12\, \mathit{eV},\ 95\text{\%}\;\mathit{C.\,L.}$ for Bayesian statistics, and ${m}_{\nu
}<2.05\, \mathit{eV},\ 95\text{\%}\;\mathit{C.\,L.}$ for the Feldman and Cousins approach. Within the present analysis , there is no statistically significant indication of a structure at the end of the spectrum. 

With deep regret we have to say that during preparation of this paper the actual leader of the experiment, the world recognized expert on neutrino mass measurements Vladimir Lobashev passed away. 

\section{Acknowledgments}
This work was supported by the Russian Foundation for Basic Research under Grants No. 93-02-039-03-a, No. 96-02-18633-a, 02-00459-a, No. 05-02-17238-a, No. 08-02-00459-a, and grant from International Science and Technology Center, No. 1076. We also would like to thank  our colleagues N. A.~Golubev, O. V.~Kazachenko, B. M.~Ovchinnikov, N. A.~Titov, Yu. I.~Zakharov, and I. E.~Yarykin for their valuable contribution to the experiment preparation and data taking. Special thanks go to Ernst Otten and Christian Weinheimer for their interest and useful discussions of the results.


\begin{thebibliography}{99}
\bibitem{oscill} K.~Nakamura et al., (Particle Data Group), J.~Phys. G {\bf 37}, 075021 (2010).
\bibitem{troitsk2003} V.M.~Lobashev, Nucl. Phys. A {\bf 719}, 153c (2003).
\bibitem{mainz} Ch.~Kraus et al., Eur. Phys. J. C {\bf 40}, 447 (2005). 
\bibitem{tkachev}F.V.~Tkachov, arXiv:physics/0604127.  
\bibitem{minuit} F.~James, CERN Program Library Long Writeup D {\bf 506}, (1994).
\bibitem{losalamos} R.G.H.~Robertson and D.~Knapp, Annu. Rev. Nucl. Part. Sci. {\bf 38}, 185 (1988).  
\bibitem{jminuit}M.~Donszelmann et al., J. Phys.: Conf. Series, {\bf 119}, 032016 (2008).
\bibitem{fermi}J.J.~Simpson, Phys. Rev. D {\bf 23}, 649 (1981).
\bibitem{states} S.~Jonsell and H.J.~Monkhorst, Phys. Pev. Lett. {\bf 76}, 4476 (1996).
\bibitem{resolution}V.N.~Aseev et al., Eur. Phys. J. D {\bf 10}, 39 (2000).
\bibitem{tkachev2}F.V.~Tkachov, arXiv:physics/0911.4271.
\bibitem{feldman} G.J.~Feldman and R.D.~Cousins, Phys. Rev. D  {\bf 57}, 3873 (1998).
\end{thebibliography}
\end{document}